\documentclass[12pt]{article}
\newcommand{\text}{\rm}
\begin{document}

\title{\textbf{Ghost condensates in Yang-Mills theories in the Landau gauge }}
\author{V.E.R. Lemes, M.S. Sarandy, and S.P. Sorella \\
{\small {\textit{UERJ - Universidade do Estado do Rio de Janeiro,}}} \\
{\small {\textit{\ Rua S\~{a}o Francisco Xavier 524, 20550-013 Maracan\~{a}, 
}}} {\small {\textit{Rio de Janeiro, Brazil.}}}}
\maketitle

\begin{abstract}
Ghost condensates of dimension two are analysed in pure $SU(N)$ Yang-Mills
theories by combining the local composite operators technique with the
algebraic BRST renormalization.
\end{abstract}

\vfill\newpage

\section{Introduction}

The understanding of the nonperturbative effects which govern the infrared
behavior of Yang-Mills theory is one of the most challenging issues in
quantum field theory. Lattice results have provided evidences of the fact
that the infrared regime of the theory is very different from the
ultraviolet one, as expected by the asymptotic freedom. Simulations of the
gluon propagator in the Landau gauge \cite{lg} and in the maximal Abelian
gauge \cite{as} display an infrared behavior compatible with the existence
of a mass gap. Attempts to explain this behavior in the continuum have been
given by taking into account Gribov's ambiguities and by using the
Schwinger-Dyson equations, see for instance refs.\cite{zw,sd} for recent
work on the subject. Nonperturbative effects are also at the origin of the
existence of the quark condensate $\left\langle \overline{q}q\right\rangle $%
, responsible for the spontaneous breaking of the chiral symmetry, and of
the pure gluon condensate $\left\langle \alpha F^{2}\right\rangle $.

Recently, much effort has been devoted to the study of condensates of
dimension two built up with gauge fields and Faddeev-Popov ghosts. For
instance, the gauge condensate $\left\langle A^{2}\right\rangle \;$has a
physical meaning in the Landau gauge \cite{gz,vb}. It plays a pivotal role
in order to account for the discrepancy recently observed between the
expected behavior of perturbation theory and the lattice results of the two
and three point functions in the Landau gauge \cite{b23, gz1}. An effective
potential for this condensate has been obtained in \cite{v1} by using the
local composite operators (LCO)\ technique \cite{v2,v3}. This result shows
that the vacuum of pure Yang-Mills theory favors a nonvanishing value of
this condensate, which gives rise to a mass term for the gluons while
contributing to the dimension four condensate $\left\langle \alpha
F^{2}\right\rangle $ through the trace anomaly. Furthermore, as shown in the
case of the compact \textit{QED }in 3\textit{D, }the condensate $%
\left\langle A^{2}\right\rangle $ turns out to have a topological meaning 
\cite{gz}, being argued to be suitable for detecting the presence of
monopole configurations. A discussion of the relevance of $\left\langle
A^{2}\right\rangle $ for confinement may be found in \cite{conf}.

In addition to the gauge condensate $\left\langle A^{2}\right\rangle $, the
ghost-antighost condensates $\left\langle \overline{c}c\right\rangle $, $%
\left\langle cc\right\rangle $ and $\left\langle \overline{c}\overline{c}%
\right\rangle $ have also received great attention \cite{ms,k,sp,dd,work} in
the context of the maximal Abelian gauge. This gauge, introduced in \cite
{th,ks}, has provided evidences for monopoles condensation as well as for
the Abelian dominance hypothesis \cite{th,abd}, which are key ingredients
for the dual superconductivity mechanism of color confinement \cite{scon,th}%
. An important point to be noted here is that the maximal Abelian gauge is
nonlinear. As a consequence, a quartic ghost interaction term must be
necessarily included for renormalizability \cite{mp,fz}. This term gives
rise to a nontrivial vacuum, corresponding to the existence of the
aforementioned ghost condensates, which display rather interesting features.
They modify the behavior of the ghost propagator in the infrared region \cite
{ms,k,sp} and lower the vacuum energy density, being interpreted as a
low-energy manifestation of the trace anomaly \cite{ms,k}. Furthermore, $%
\left\langle \overline{c}c\right\rangle $ is believed to be part of a more
general dimension two condensate, namely $\left( \frac{1}{2}\left\langle
AA\right\rangle -\xi \left\langle c\overline{c}\right\rangle \right) $,
where $\xi $ denotes the gauge parameter of the maximal Abelian gauge. This
condensate has been proposed in \cite{ope} due to the property of being
on-shell BRST\ invariant, and it is expected to provide effective masses for
both off-diagonal gauge and ghost fields \cite{ope,dd}, thus playing an
important role for the Abelian dominance. It is worth remarking that the
ghost condensates do not show up only in the maximal Abelian gauge, being
present indeed in other nonlinear gauges as, for instance the so called
Curci-Ferrari gauge \cite{cf,cf1,cf2}. Also, the effective potential for the
ghost-gluon condensate $\left( \frac{1}{2}\left\langle AA\right\rangle -\xi
\left\langle c\overline{c}\right\rangle \right) $ in this gauge has recently
been obtained  \cite{gg}.

The aim of this work is to analyze the ghost condensates in the Landau
gauge. Although in this gauge there is no quartic ghost interaction, these
condensates can be studied by means of the LCO technique, as improved by 
\cite{v2,v3}. In particular, by combining the LCO\ technique with the BRST\
algebraic renormalization \cite{book}, we shall be able to obtain the
effective potential for $\left\langle cc\right\rangle $ and $\left\langle 
\overline{c}\overline{c}\right\rangle $. This result may signal a deeper
physical meaning of these condensates.

The LCO formalism allows to obtain in a consistent way the effective
potential for local composite operators. The method consists of introducing
in the starting action the set of local composite operators to be studied
coupled to suitable external sources. However, one has to be sure that all
needed counterterms have been included. This requires the introduction of
counterterms quadratic in the external sources. As shown in \cite{v2,v3},
these quadratic terms can be dealt with by introducing a
Hubbard-Stratonovich transformation allowing to obtain a renormalizable
effective potential which is linear in the external sources, being thus
compatible with the density energy interpretation of the effective
potential. The method has already been successfully applied to the
Gross-Neveu model \cite{v2} and to the $\lambda \phi ^{4}$ and
Coleman-Weinberg models \cite{v3}. In the latter case, a manifest gauge
invariant two-loop effective potential has been obtained, according to the
Nielsen identities.

It is worth underlining here that the LCO\ technique can be combined with
the BRST\ algebraic renormalization \cite{book} in order to obtain all
needed counterterms. The algebraic renormalization allows indeed to write
down a suitable set of Ward identities which determine, to all orders of
perturbation theory, the most general counterterm depending on the fields
and external sources needed for the renormalizability of a given set of
local composite operators.

The paper is organized as follows. In Sect.2 we give a brief review of the
LCO technique in the case of the condensate $\left\langle A^{2}\right\rangle 
$ for Yang-Mills in the Landau gauge. This example will allow us to show how
the algebraic renormalization can be combined with the LCO method to provide
a very powerful framework for the analysis of the effective potential.
Sect.3 is devoted to the study of the ghost condensates $\left\langle
cc\right\rangle $ and $\left\langle \overline{c}\overline{c}\right\rangle .\;
$In Sect.4 we present the conclusions.

\section{The LCO technique for the gauge condensate $\left\langle A_{\mu
}^{a}A^{\mu a}\right\rangle \;$}

Let us begin with a brief review of the LCO technique \cite{v2,v3} in the
case of the evaluation of the effective potential for the gauge condensate $%
\left\langle A_{\mu }^{a}A^{\mu a}\right\rangle $ in the Landau gauge \cite
{v1}. It is known that we can give a meaning to the local composite operator 
$A_{\mu }^{a}A^{\mu a}$ in the Landau gauge. Indeed, due to the
transversality condition $\partial _{\mu }A^{\mu a}=0$, the integrated mass
dimension two operator $\int d^{4}xA_{\mu }^{a}A^{\mu a}$ is gauge
invariant. It makes sense therefore to look at the condensate $\left\langle
A_{\mu }^{a}A^{\mu a}\right\rangle $ in the Landau gauge.

In order to evaluate the effective potential for $A_{\mu }^{a}A^{\mu a}$ we
introduce it in the starting action by means of an external source $j(x)$ of
ghost number zero and dimension two \cite{v1}, namely

\begin{equation}
S_{\mathrm{YM}}+S_{\mathrm{gf}}+\int d^{4}x\frac{1}{2}jA_{\mu }^{a}A^{\mu
a}\;,  \label{ai}
\end{equation}
where

\begin{eqnarray}
S_{\mathrm{YM}} &=&-\frac{1}{4}\int d^{4}xF^{a\mu \nu }F_{\mu \nu }^{a}\;,
\label{sym} \\
S_{\mathrm{gf}} &=&\int d^{4}x\left( b^{a}\partial _{\mu }A^{\mu a}+%
\overline{c}^{a}\partial ^{\mu }\left( D_{\mu }c\right) ^{a}\right) \;, 
\nonumber
\end{eqnarray}
and

\begin{eqnarray}
F_{\mu \nu }^{a} &=&\partial _{\mu }A_{\nu }^{a}-\partial _{\nu }A_{\mu
}^{a}+gf^{abc}A_{\mu }^{b}A_{\nu }^{c}\;,  \label{f} \\
\left( D_{\mu }c\right) ^{a} &=&\partial _{\mu }c^{a}+gf^{abc}A_{\mu
}^{b}c^{c}\;.  \nonumber
\end{eqnarray}
As it is well known, the gauge fixed action $\left( S_{\mathrm{YM}}+S_{%
\mathrm{gf}}\right) $ is left invariant by the BRST nilpotent transformations

\begin{eqnarray}
sA_{\mu }^{a} &=&-\left( D_{\mu }c\right) ^{a}\;,\;\;\;\;\;\;sc^{a}=\frac{g}{%
2}f^{abc}c^{b}c^{c}\;,  \label{brst} \\
s\overline{c}^{a} &=&b^{a}\;\;\;\;\;\;\;,\;\;\;\;\;\;sb^{a}=0\;.  \nonumber
\end{eqnarray}
However, before starting the computation, one has to be sure that the action 
$\left( \ref{ai}\right) $ contains all needed counterterms. In other words
we have to start with the most general action compatible with the
invariances of the model and with the power-counting. Observe also that
expression $\left( \ref{ai}\right) $ with the inclusion of the coupling $%
jA^{2}$ is renormalizable by power-counting. The search for the most general
action is easily handled by using the algebraic renormalization \cite{book}.
This will be the task of the next section.

\subsection{BRST\ characterization of the most general counterterm}

In order to characterize the most general action, including all needed
counterterms, we proceed by introducing a BRST\ doublet \cite{book,bbh} of
external sources $\left( j(x),\lambda (x)\right) $, according to 
\begin{equation}
s\lambda =j\;,\;\;\;\;\;\;sj=0\;,  \label{jld}
\end{equation}
where $\lambda $ is an external source of dimension one and ghost number -1.

\[
\begin{tabular}{|l|l|l|}
\hline
& $j$ & $\lambda $ \\ \hline
Gh. number & \thinspace 0 & \thinspace \thinspace -1 \\ \hline
Dimension & \thinspace 2 & $\,\,$1 \\ \hline
\end{tabular}
\]
It is apparent then that the quantity

\begin{equation}
S_{\mathrm{LCO}}=s\int d^{4}x\left( \lambda \frac{A^{2}}{2}+\frac{\varsigma 
}{2}\lambda j\right) =\int d^{4}x\left( \frac{1}{2}jA_{\mu }^{a}A^{\mu
a}+\lambda A_{\mu }^{a}\partial ^{\mu }c^{a}+\frac{\varsigma }{2}%
j^{2}\right) \;,  \label{lj}
\end{equation}
is BRST invariant. The meaning of the external source $\lambda (x)$ is also
clear; it defines the composite operator $A_{\mu }^{a}\partial ^{\mu }c^{a}$
which is the BRST variation of $A^{2}/2$.\ Observe also that the BRST\
invariance and the power-counting naturally allow for the term $%
j^{2}=s(\lambda j)$. According now to \cite{v1,v2,v3}, the parameter $%
\varsigma $ will be fixed by demanding that it depends on the gauge
coupling, namely $\varsigma =\varsigma (g)$. The dependence of $\varsigma $
from $g$ can be computed order by order in the loop expansion

\begin{equation}
\varsigma (g)=\varsigma _{0}+\hbar \varsigma _{1}+\hbar ^{2}\varsigma
_{2}+...  \label{ls}
\end{equation}
and is obtained from the renormalization group equations \cite{v1,v2,v3}.
This is a highly nontrivial requirement, which enables the LCO technique to
capture nonperturbative effects \cite{v1,v2,v3}. In particular, the lowest
order coefficient $\varsigma _{0}$ plays a fundamental role in order to
obtain a nontrivial vacuum configuration for $\left\langle
A^{2}\right\rangle $. In fact, as shown in \cite{v1}, this coefficient is
scheme independent\footnote{%
We are indebted to D. Dudal for many valuable discussions on this important
point.} and carries nonperturbative information, allowing for a nonvanishing
value of $\left\langle A^{2}\right\rangle $.

Let us turn now to write down the Ward identities which will ensure the
renormalizability of the model. We begin by observing that the action $%
\left( S_{\mathrm{YM}}+S_{\mathrm{gf}}+S_{\mathrm{LCO}}\right) $ is
invariant under the BRST$\;$transformations\ $\left( \ref{brst}\right)
,\left( \ref{jld}\right) $%
\begin{equation}
s\left( S_{\mathrm{YM}}+S_{\mathrm{gf}}+S_{\mathrm{LCO}}\right) =0\;.
\label{a}
\end{equation}
Introducing also the external BRST\ invariant sources $\left( \Omega _{\mu
}^{a},L^{a}\right) $ coupled to the nonlinear variations of the fields $%
A_{\mu }^{a}\;$and $c^{a}$ \cite{book}

\begin{equation}
S_{\mathrm{ext}}=\int d^{4}x\left( -\Omega ^{a\mu }\left( D_{\mu }c\right)
^{a}+L^{a}\frac{g}{2}f^{abc}c^{b}c^{c}\right) \;,  \label{sext}
\end{equation}
it turns out that the complete action 
\begin{equation}
\Sigma =S_{\mathrm{YM}}+S_{\mathrm{gf}}+S_{\mathrm{LCO}}+S_{\mathrm{ext}}\;,
\label{ca}
\end{equation}
obeys the following identities \cite{book}

\begin{itemize}
\item  the Slavnov-Taylor identity 
\begin{eqnarray}
\mathcal{S}(\Sigma ) &=&0\;,  \label{st} \\
\mathcal{S}(\Sigma ) &=&\int d^{4}x\left( \frac{\delta \Sigma }{\delta
A_{\mu }^{a}}\frac{\delta \Sigma }{\delta \Omega ^{a\mu }}+\frac{\delta
\Sigma }{\delta c^{a}}\frac{\delta \Sigma }{\delta L^{a}}+b^{a}\frac{\delta
\Sigma }{\delta \overline{c}^{a}}+j\frac{\delta \Sigma }{\delta \lambda }%
\right) \;,  \nonumber
\end{eqnarray}

\item  the Landau gauge condition and the antighost equation 
\begin{equation}
\frac{\delta \Sigma }{\delta b^{a}}=\partial _{\mu }A^{\mu a}\;,\;\;\;\;\;\;%
\frac{\delta \Sigma }{\delta \overline{c}^{a}}+\partial ^{\mu }\frac{\delta
\Sigma }{\delta \Omega ^{a\mu }}=0\;,  \label{la}
\end{equation}

\item  the linearly broken integrated ghost equation$\;$Ward identity 
\begin{equation}
\int d^{4}x\left( \frac{\delta \Sigma }{\delta c^{a}}+gf^{abc}\overline{c}%
^{b}\frac{\delta \Sigma }{\delta b^{c}}\right) =\Delta _{\mathrm{cl}}^{a}\;,
\label{gi}
\end{equation}
where $\Delta _{\mathrm{cl}}^{a}$ is the classical breaking 
\begin{equation}
\Delta _{\mathrm{cl}}^{a}=\int d^{4}x\left( gf^{abc}\Omega ^{b\mu }A_{\mu
}^{c}-gf^{abc}L^{b}c^{c}\right) \;.  \label{cb}
\end{equation}
We underline that the ghost equation Ward identity $\left( \ref{gi}\right) $
is an important feature of the Landau gauge, which has allowed for an
algebraic proof of several nonrenormalization results \cite{landau}.
\end{itemize}

Let us now look at the most general counterterm depending on the external
sources $\left( j,\lambda \right) .$ Taking into account that $\left(
j,\lambda \right) $ form a BRST\ doublet, it turns out that the most general
expression depending on $\left( j,\lambda \right) $ with ghost number zero
and dimension four is 
\begin{eqnarray}
S_{\mathrm{LCO}}^{\mathrm{count}} &=&s\int d^{4}x\left( a_{1}\lambda \frac{%
A^{2}}{2}+\frac{a_{2}}{2}\lambda j+a_{3}\lambda \overline{c}^{a}c^{a}\right)
\label{ct} \\
&=&\int d^{4}x\left( a_{1}j\frac{A^{2}}{2}+a_{1}\lambda A_{\mu }^{a}\partial
^{\mu }c^{a}+\frac{a_{2}}{2}j^{2}+a_{3}j\overline{c}^{a}c^{a}\;\right. 
\nonumber \\
&&\;\;\;\;\;\;\;\;\left. -a_{3}\lambda b^{a}c^{a}+\frac{a_{3}}{2}\lambda 
\overline{c}^{a}f^{abc}c^{b}c^{c}\right) \;.  \nonumber
\end{eqnarray}
However, $a_{3}=0\;$due to the Landau ghost equation Ward identity $\left( 
\ref{gi}\right) $ and to the gauge fixing condition and antighost equation $%
\left( \ref{la}\right) $. Thus

\begin{equation}
S_{\mathrm{LCO}}^{\mathrm{count}}=\int d^{4}x\left( a_{1}j\frac{A^{2}}{2}%
+a_{1}\lambda A_{\mu }^{a}\partial ^{\mu }c^{a}+\frac{a_{2}}{2}j^{2}\right)
\;,  \label{ctf}
\end{equation}
which has precisely the same form as $S_{\mathrm{LCO}}$. The coefficient $%
a_{1}$ is related to the renormalization of the source $j$ and therefore to
the renormalization of the local operator $A^{2}.$ The coefficient $a_{2}$
can be associated to a renormalization of the parameter $\varsigma $. Its
physical meaning is that of taking into account the divergences which appear
in the correlator $\left\langle A^{2}(x)A^{2}(y)\right\rangle \;$\cite{v1}.
Finally we observe that, as expected by BRST invariance, the renormalization
of the operator $A_{\mu }^{a}\partial ^{\mu }c^{a}$ is related to that of $%
A^{2}$. Having characterized the most general local counterterm depending on
the external sources $\left( j,\lambda \right) $, we are now ready to
evaluate the one-loop effective potential for the condensate $\left\langle
A^{2}\right\rangle $.

\subsection{The effective potential for $\left\langle A^{2}\right\rangle $}

The first step in order to study the condensate $\left\langle
A^{2}\right\rangle $ is to analyse the generating functional $\mathcal{W}%
(j).\;$Setting thus to zero the external sources $\Omega _{\mu
}^{a},L^{a},\lambda ,\;$we have

\begin{equation}
\exp i\mathcal{W}(j)=\int \left[ D\varphi \right] \exp i\left( S_{\mathrm{YM}%
}+S_{\mathrm{gf}}+\int d^{4}x\left( \frac{1}{2}jA_{\mu }^{a}A^{\mu a}\;+%
\frac{\varsigma }{2}j^{2}\right) \right) \;,  \label{gf}
\end{equation}
where $\left[ D\varphi \right] $ denotes integration over all quantum fields 
$(A_{\mu }^{a},b^{a},\overline{c}^{a},c^{a})$. Taking the functional
derivative of expression $\left( \ref{gf}\right) $ we obtain

\begin{equation}
\left. \frac{\delta \mathcal{W}(j)}{\delta j}\right| _{j=0}=\frac{1}{2}%
\left\langle A_{\mu }^{a}A^{\mu a}\right\rangle \;.  \label{djw}
\end{equation}
In order to deal with the term $j^{2}$ we follow \cite{v1}, introducing a
Hubbard-Stratonovich field $\sigma $ so that

\begin{equation}
\left( \frac{1}{2}jA_{\mu }^{a}A^{\mu a}\;+\frac{\varsigma }{2}j^{2}\right) =%
\frac{1}{2}\left( \sqrt{\varsigma }j+\frac{1}{2\sqrt{\varsigma }}A_{\mu
}^{a}A^{\mu a}\right) ^{2}-\frac{1}{8\varsigma }\left( A_{\mu }^{a}A^{\mu
a}\right) ^{2}\;,  \label{f1}
\end{equation}
and

\begin{equation}
\frac{1}{2}\left( \sqrt{\varsigma }j+\frac{1}{2\sqrt{\varsigma }}A_{\mu
}^{a}A^{\mu a}\right) ^{2}\Rightarrow -\frac{\sigma ^{2}}{2g^{2}}+\frac{1}{g}%
\sigma \left( \sqrt{\varsigma }j+\frac{1}{2\sqrt{\varsigma }}A_{\mu
}^{a}A^{\mu a}\right) \;.  \label{f2}
\end{equation}
As a consequence, for the functional generator $\mathcal{W}(j)$ we get

\begin{equation}
\exp i\mathcal{W}(j)=\int \left[ D\varphi \right] D\sigma \exp i\left(
S(A,\sigma )+\int d^{4}x\frac{\sqrt{\varsigma }}{g}\sigma j\right) \;,
\label{wjsa}
\end{equation}
with

\begin{equation}
S(A,\sigma )=S_{\mathrm{YM}}+S_{\mathrm{gf}}+\int d^{4}x\left( -\frac{\sigma
^{2}}{2g^{2}}+\frac{\sigma }{2g\sqrt{\varsigma }}A_{\mu }^{a}A^{\mu a}-\frac{%
1}{8\varsigma }\left( A_{\mu }^{a}A^{\mu a}\right) ^{2}\right) \;.
\label{sas}
\end{equation}
In particular,

\begin{equation}
\left. \frac{\delta \mathcal{W}(j)}{\delta j}\right| _{j=0}=\frac{1}{2}%
\left\langle A_{\mu }^{a}A^{\mu a}\right\rangle \;=\frac{\sqrt{\varsigma }}{g%
}\left\langle \sigma \right\rangle _{S(A,\sigma )}\;.  \label{ff}
\end{equation}
This is a very remarkable identity \cite{v1}, stating that the condensate $%
\left\langle A^{2}\right\rangle \;$is related to the nonvanishing
expectation value of $\sigma $ evaluated with the new action $S(A,\sigma )$.
Let us proceed thus with the computation of the effective potential for $%
\sigma $. We shall limit here to the one-loop order, which already displays
all the nonperturbative features of the LCO technique.

\subsection{Evaluation of the one-loop effective potential}

In order to compute the effective potential for $\sigma $ at the one-loop
order only the quadratic part of the action $S(A,\sigma )$ is relevant,
namely

\begin{equation}
S^{\mathrm{quad}}(A,\sigma )=-\frac{\sigma ^{2}}{2g^{2}}\int d^{4}x+\frac{1}{%
2}\int d^{4}xA^{a\mu }\mathcal{M}_{\mu \nu }^{ab}A^{b\nu }\;,  \label{sq}
\end{equation}
with

\begin{equation}
\mathcal{M}_{\mu \nu }^{ab}=\left[ \left( \partial ^{2}g_{\mu \nu }-\left( 1-%
\frac{1}{\alpha }\right) \partial _{\mu }\partial _{\nu }\right) +\frac{%
\sigma }{g\sqrt{\varsigma }}g_{\mu \nu }\right] \delta ^{ab}=\mathcal{A}%
_{\mu \nu }\delta ^{ab}\;,  \label{m}
\end{equation}
where $\sigma $ is now a constant field and the limit $\alpha \rightarrow 0$
has to be taken at the end in order to implement the Landau gauge. For the
one-loop effective potential we have

\begin{equation}
V^{\mathrm{eff}}(\sigma )=\frac{\sigma ^{2}}{2g^{2}}-\frac{i}{2}\mathrm{%
tr\log \det }\mathcal{M}_{\mu \nu }^{ab}\;.  \label{p}
\end{equation}
From

\begin{equation}
\mathrm{\det }\mathcal{M}_{\mu \nu }^{ab}=\left( \mathrm{\det }\mathcal{A}%
_{\mu \nu }\right) ^{\left( N^{2}-1\right) }\;,  \label{ma}
\end{equation}
we have

\begin{equation}
V^{\mathrm{eff}}(\sigma )=\frac{\sigma ^{2}}{2g^{2}}-\frac{i\left(
N^{2}-1\right) }{2}\mathrm{tr\log \det }\mathcal{A}_{\mu \nu }\;.  \label{p1}
\end{equation}
Making use of the dimensional regularization and recalling that

\begin{equation}
\left. \mathrm{\log \det }\mathcal{A}_{\mu \nu }\right| _{\alpha \rightarrow
0}=(d-1)\log \left( \partial ^{2}+\frac{\sigma }{g\sqrt{\varsigma }}\right) +%
\mathrm{const.}  \label{p2}
\end{equation}
it follows

\begin{equation}
V^{\mathrm{eff}}(\sigma )=\frac{\sigma ^{2}}{2g^{2}}-\frac{1}{2}\left(
N^{2}-1\right) i\int \frac{d^{d}k}{\left( 2\pi \right) ^{d}}(d-1)\log \left(
-k^{2}+\frac{\sigma }{g\sqrt{\varsigma }}\right) \;.  \label{p3}
\end{equation}
From

\begin{equation}
i\int \frac{d^{d}k}{\left( 2\pi \right) ^{d}}\log \left( \frac{\sigma }{g%
\sqrt{\varsigma }}-k^{2}\right) =\frac{1}{2\left( 4\pi \right) ^{2}}\left( 
\frac{\sigma }{g\sqrt{\varsigma }}\right) ^{2}\left[ \frac{1}{\varepsilon }%
-\gamma +\frac{3}{2}-\log \frac{\sigma }{g\sqrt{\varsigma }4\pi \mu ^{2}}%
\right] \;\;  \label{dr}
\end{equation}
the one-loop effective potential in the $\overline{MS}$ scheme is found to be

\begin{equation}
\;V^{\mathrm{eff}}(\sigma )=\frac{\sigma ^{2}}{2g^{2}}+\hbar \frac{3\left(
N^{2}-1\right) }{64\pi ^{2}}\left( \frac{1}{g^{2}\varsigma }\right) \sigma
^{2}\left( \log \frac{\sigma }{g\sqrt{\varsigma }\overline{\mu }^{2}}-\frac{5%
}{6}\right) ,  \label{pf}
\end{equation}
where the presence of the loop expansion parameter $\hbar $ has been made
explicit. Let us now proceed with the evaluation of the first term of the
series $\left( \ref{ls}\right) $. To this purpose we require that the
effective potential $V^{\mathrm{eff}}(\sigma )$ obeys the renormalization
group equation, namely 
\begin{equation}
\mu \frac{dV^{\mathrm{eff}}(\sigma )}{d\mu }=0+O(\hbar ^{2})\;.  \label{aeq}
\end{equation}
To work out this condition the knowledge of the running of the field $\sigma 
$ is needed. This is easily achieved with the help of the relation between $%
\sigma $ and $\left\langle A^{2}\right\rangle $ of eq.$\left( \ref{ff}%
\right) $, namely 
\begin{equation}
\left\langle \frac{A^{2}}{2}\right\rangle =\frac{\sqrt{\varsigma }}{g}\sigma
\;.  \label{ffr}
\end{equation}
A simple calculation gives 
\begin{equation}
\mu \partial _{\mu }\sigma =\hbar \left( \gamma _{A^{2}}^{(1)}+\frac{\beta
_{g}^{(1)}}{g}-\frac{1}{2}\frac{\beta _{g}^{(1)}}{\varsigma _{0}}\frac{%
\partial \varsigma _{0}}{\partial g}\right) \sigma +O(\hbar ^{2})\;,
\label{mds}
\end{equation}
where $\gamma _{A^{2}}^{(1)}$ is the one-loop anomalous dimension of the
local operator $A^{2}$ and, since $\varsigma =\varsigma (g)$, 
\begin{equation}
\mu \partial _{\mu }\varsigma =\beta _{g}\frac{\partial \varsigma }{\partial
g}=\hbar \beta _{g}^{(1)}\frac{\partial \varsigma _{0}}{\partial g}+O(\hbar
^{2})\;.  \label{mdxi}
\end{equation}
>From eq.$\left( \ref{aeq}\right) $ it follows that

\begin{equation}
\frac{1}{\varsigma _{0}}=\frac{16\pi ^{2}}{3\left( N^{2}-1\right) }\left(
2\gamma _{A^{2}}^{(1)}+2\frac{\beta _{g}^{(1)}}{g}\right) \;,  \label{ffxi0}
\end{equation}
from which one sees that $\varsigma _{0}$ is scheme independent. The
explicit value of $\gamma _{A^{2}}^{(1)}$ can be found in \cite{v1,gr,dsv}
and turns out to be

\begin{equation}
\gamma _{A^{2}}^{(1)}=\frac{g^{2}N}{16\pi ^{2}}\frac{35}{6}\;.  \label{ga}
\end{equation}
Therefore, recalling that 
\begin{equation}
\beta _{g}^{(1)}=-\frac{11N}{3}\frac{g^{3}}{16\pi ^{2}}\;,  \label{betasun}
\end{equation}
for $\varsigma _{0}$ one gets

\begin{equation}
\varsigma _{0}=\frac{1}{g^{2}}\frac{9}{13}\frac{\left( N^{2}-1\right) }{N}\;,
\label{lc}
\end{equation}
in complete agreement with \cite{v1}. This expression displays the
nonperturbative features of the LCO method \cite{v1}, emphasized by the
presence in eq.$\left( \ref{lc}\right) $ of the factor $1/\left(
g^{2}N\right) .$

\section{The ghost condensates}

Having given a brief summary of the LCO technique for the evaluation of $%
\left\langle A^{2}\right\rangle $, let us turn to the study of the ghost
condensates. We shall consider the following local composite operators

\begin{equation}
f^{abc}c^{b}c^{c}\;\;\;\mathrm{and\;\;\;\;\;}f^{abc}\overline{c}^{b}%
\overline{c}^{c}\;.  \label{go}
\end{equation}
carrying ghost number $2$ and $-2,$ respectively.

Following now the LCO procedure, we first introduce the two operators $%
f^{abc}c^{b}c^{c},\mathrm{\;}f^{abc}\overline{c}^{b}\overline{c}^{c}\;$in
the starting action by means of a set of external sources $\left(
L^{a},\;\tau ^{a},\;\eta ^{a}\right) $ transforming as

\begin{eqnarray}
s\eta ^{a} &=&\tau ^{a}\;,  \label{ss} \\
s\tau ^{a} &=&0\;,  \nonumber \\
sL^{a} &=&0\;.  \nonumber
\end{eqnarray}
$\;$The quantum numbers of $\left( L^{a},\;\tau ^{a},\;\eta ^{a}\right) $
are displayed in the following table

\begin{equation}
\begin{tabular}{|l|l|l|l|}
\hline
& $L^{a}$ & $\tau ^{a}$ & $\eta ^{a}$ \\ \hline
Gh. number & $-2$ & $2$ & $1$ \\ \hline
Dimension & $2$ & $2$ & $1$ \\ \hline
\end{tabular}
\label{qltn}
\end{equation}
Notice also that the sources $\left( \tau ^{a},\;\eta ^{a}\right) $ form a
BRST doublet. Moreover, according to the algebraic renormalization, the
BRST\ invariant source $L^{a}$ is needed to define the nonlinear variation
of the ghost $c^{a}$, which is $\left( gf^{abc}c^{b}c^{c}\right) /2$. In the
present case, the BRST\ invariant part of the action depending on the
external sources $\left( L^{a},\;\tau ^{a},\;\eta ^{a}\right) $ reads

\begin{eqnarray}
S_{\mathrm{LCO}} &=&s\int d^{4}x\left( L^{a}c^{a}\;-\frac{g}{2}\eta
^{a}f^{abc}\overline{c}^{b}\overline{c}^{c}+\varsigma \eta ^{a}L^{a}\right)
\,  \label{lcogh} \\
&=&\int d^{4}x\left( L^{a}\frac{g}{2}f^{abc}c^{b}c^{c}\;-\frac{g}{2}\tau
^{a}f^{abc}\overline{c}^{b}\overline{c}^{c}+g\eta ^{a}f^{abc}b^{b}\overline{c%
}^{c}+\varsigma \tau ^{a}L^{a}\right) \;,  \nonumber
\end{eqnarray}
where, as before, $\varsigma $ is a dimensionless parameter which will be
fixed later on by demanding the compatibility with the renormalization group
equations. Notice also that the presence of the term $\tau ^{a}L^{a}$ is
allowed by the power counting. For the complete action $\Sigma $ we have now

\begin{equation}
\Sigma =S_{\mathrm{YM}}+S_{\mathrm{m}}+S_{\mathrm{gf}}+S_{\mathrm{LCO}}+S_{%
\mathrm{ext}}\;,  \label{cagh}
\end{equation}
where $S_{\mathrm{m}}$ stands for the fermionic matter action, which we
include for completeness

\begin{equation}
S_{\mathrm{m}}=\int d^{4}x\overline{\psi }^{iI}i\gamma ^{\mu }(\partial
_{\mu }\delta ^{IJ}-igA_{\mu }^{a}T^{aIJ})\psi ^{iJ}\;,  \label{mt}
\end{equation}
where $T^{aIJ}$ are the generators of $SU(N)$ in the fundamental
representation, and $1\leq i\leq n_{f}$ gives the number of fermions. Also,
for the BRST transformations of the matter fields we have

\begin{eqnarray}
s\psi ^{iI} &=&-igc^{a}T^{aIJ}\psi ^{iJ}\;,  \label{spsi} \\
s\overline{\psi }^{iI} &=&-ig\overline{\psi }^{iJ}T^{aJI}c^{a}\;,  \nonumber
\\
sS_{\mathrm{m}} &=&0\;.  \nonumber
\end{eqnarray}
Finally, introducing external sources $\Omega ^{a\mu },\overline{Y}%
^{iI},\;Y^{iI},\;$ coupled to the nonlinear variations of the fields $A_{\mu
}^{a},$ $\psi ^{iI}$ and $\overline{\psi }^{iI}$, the external part of the
complete action $\left( \ref{cagh}\right) $ reads

\begin{equation}
S_{\mathrm{ext}}=\int d^{4}x\,\left[ \Omega ^{a\mu }\left( sA_{\mu
}^{a}\right) \;+\overline{Y}^{iI}\left( s\psi ^{iI}\right) +\left( s%
\overline{\psi }^{iI}\right) Y^{iI}\right] \;.  \label{extgh}
\end{equation}
It turns out that $\Sigma $ is constrained by the following identities:

\begin{itemize}
\item  the Slavnov-Taylor identity 
\begin{equation}
\mathcal{S}(\Sigma )=0\;,  \label{stf}
\end{equation}
\begin{eqnarray}
\mathcal{S}(\Sigma ) &=&\int d^{4}x\left[ \frac{\delta \Sigma }{\delta
A_{\mu }^{a}}\frac{\delta \Sigma }{\delta \Omega ^{a\mu }}+\frac{\delta
\Sigma }{\delta \overline{\psi }^{iI}}\frac{\delta \Sigma }{\delta Y^{iI}}+%
\frac{\delta \Sigma }{\delta \overline{Y}^{iI}}\frac{\delta \Sigma }{\delta
\psi ^{iI}}\right.   \nonumber \\
&&\left. \;\;\;\;\;+\left( \frac{\delta \Sigma }{\delta L^{a}}-\varsigma
\tau ^{a}\right) \frac{\delta \Sigma }{\delta c^{a}}+b^{a}\frac{\delta
\Sigma }{\delta \overline{c}^{a}}+\tau ^{a}\frac{\delta \Sigma }{\delta \eta
^{a}}\right] \;.  \label{stfop}
\end{eqnarray}

\item  the Landau gauge condition and the antighost equation 
\begin{equation}
\frac{\delta \Sigma }{\delta b^{a}}=\partial _{\mu }A^{\mu a}-gf^{abc}\eta
^{b}\overline{c}^{c}\;,\;\;\;\;\;\;  \label{lgh}
\end{equation}
\begin{equation}
\frac{\delta \Sigma }{\delta \overline{c}^{a}}+\partial ^{\mu }\frac{\delta
\Sigma }{\delta \Omega ^{a\mu }}=gf^{abc}\tau ^{b}\overline{c}%
^{c}-gf^{abc}\eta ^{b}b^{c}\;.  \label{agh}
\end{equation}
Notice that both the Landau gauge condition and the antighost equation gets
modified by terms which are linear in the quantum fields, representing
therefore classical breakings.

\item  the modified linearly broken integrated ghost equation$\;$Ward
identity 
\begin{equation}
\int d^{4}x\left( \frac{\delta \Sigma }{\delta c^{a}}+gf^{abc}\overline{c}%
^{b}\frac{\delta \Sigma }{\delta b^{c}}+gf^{abc}\eta ^{b}\frac{\delta \Sigma 
}{\delta \tau ^{c}}\right) =\Delta _{\mathrm{cl}}^{a}\;,  \label{wgh}
\end{equation}
where $\Delta _{\mathrm{cl}}^{a}$ is the classical breaking 
\begin{equation}
\Delta _{\mathrm{cl}}^{a}=g\int d^{4}x\left[ f^{abc}\left( \Omega ^{b\mu
}A_{\mu }^{c}-L^{b}c^{c}+\varsigma \eta ^{b}L^{c}\right) -i\overline{Y}%
^{i}T^{a}\psi ^{i}+i\overline{\psi }^{i}T^{a}Y^{i}\;\right]   \label{bgh}
\end{equation}
\end{itemize}

Repeating now the same procedure as before, it is not difficult to prove
that the complete action $\Sigma $ is stable with respect to radiative
corrections. In other words $\Sigma $ is the most general action compatible
with the Ward identities defining the model, up to a multiplicative
renormalization of the fields, sources, the gauge coupling $g$ and the
parameter$\,\varsigma $. In particular, concerning the allowed counterterms
depending on the external sources $\left( L^{a},\tau ^{a},\eta ^{a}\right) $%
, it follows that the antighost equation $\left( \ref{agh}\right) $ and the
ghost Ward identity $\left( \ref{wgh}\right) $ imply  the absence of
counterterms of the type $L^{a}f^{abc}c^{b}c^{c},$ $\;\tau ^{a}f^{abc}%
\overline{c}^{b}\overline{c}^{c}$ and $\eta ^{a}f^{abc}b^{b}\overline{c}^{c}$
\cite{landau,book,dsv}. The only allowed counterterm in the external sources 
$\left( L^{a},\tau ^{a},\eta ^{a}\right) $ is given by $\tau ^{a}L^{a}$.
This counterterm corresponds to the renormalization of the LCO parameter $%
\varsigma $, stemming from the logarithmic divergences of the vacuum
correlator $\left\langle c^{2}(x)\overline{c}^{2}(y)\right\rangle $.

\subsection{The effective potential for $\left\langle
f^{abc}c^{b}c^{c}\right\rangle $ and $\left\langle f^{abc}\overline{c}^{b}%
\overline{c}^{c}\right\rangle $}

We can now begin the discussion of the effective potential for the local
ghost operators $f^{abc}c^{b}c^{c}$ and $f^{abc}\overline{c}^{b}\overline{c}%
^{c}$. Let us start with the generating functional $\mathcal{W}(L,\tau ).\;$%
Setting thus to zero the external sources $\Omega _{\mu }^{a},\eta ^{a},\;%
\overline{Y}^{iI},Y^{iI}$ we have

\begin{eqnarray}
\exp i\mathcal{W}(L,\tau ) &=&\int \left[ D\varphi \right] \exp i\left( S_{%
\mathrm{YM}}+S_{\mathrm{m}}+S_{\mathrm{gf}}\right)   \label{gfgh} \\
&&\times \exp i\left[ \int d^{4}x\left( L^{a}\frac{gf^{abc}c^{b}c^{c}}{2}%
\;-\tau ^{a}\frac{gf^{abc}\overline{c}^{b}\overline{c}^{c}}{2}\,+\varsigma
\tau ^{a}L^{a}\right) \right] \;  \nonumber
\end{eqnarray}
where $\left[ D\varphi \right] $ denotes integration over all quantum fields 
$(A_{\mu }^{a},b^{a},\overline{c}^{a},c^{a},\psi ^{iJ},\overline{\psi }^{iJ})
$. Taking the functional derivative of expression $\left( \ref{gfgh}\right) $
we obtain

\begin{equation}
\left. \frac{\delta \mathcal{W}(L,\tau )}{\delta L^{a}}\right| _{L,\tau
=0}=\left\langle \frac{g}{2}f^{abc}c^{b}c^{c}\right\rangle \;,  \label{dl}
\end{equation}
and

\begin{equation}
\left. \frac{\delta \mathcal{W}(L,\tau )}{\delta \tau ^{a}}\right| _{L,\tau
=0}=-\left\langle \frac{g}{2}f^{abc}\overline{c}^{b}\overline{c}%
^{c}\,\right\rangle \;.  \label{dt}
\end{equation}
In order to deal with the quadratic term $\tau ^{a}L^{a}$ we introduce a
pair of Hubbard-Stratonovich fields $\left( \sigma ^{a},\overline{\sigma }%
^{a}\,\right) \;$in the adjoint representation

\begin{equation}
\begin{tabular}{|l|l|l|}
\hline
& $\sigma ^{a}$ & $\overline{\sigma }^{a}\,$ \\ \hline
Gh. number & $2$ & $-2$ \\ \hline
Dimension & $2$ & $2$ \\ \hline
\end{tabular}
\label{hb}
\end{equation}
so that
\begin{equation}
L^{a}\frac{g}{2}f^{abc}c^{b}c^{c}\;-\frac{g}{2}\tau ^{a}f^{abc}\overline{c}%
^{b}\overline{c}^{c}\,+\varsigma \tau ^{a}L^{a}\,\Rightarrow \mathcal{L}%
_{\sigma \overline{\sigma }}  \label{hsgh}
\end{equation}
where
\begin{eqnarray}
&&\mathcal{L}_{\sigma \overline{\sigma }}=-\frac{1}{g^{2}}\sigma ^{a}%
\overline{\sigma }^{a}\,+\frac{\overline{\sigma }^{a}}{g\sqrt{\varsigma }}%
\left( \frac{g}{2}f^{abc}c^{b}c^{c}\right) +\frac{\sqrt{\varsigma }}{g}%
\overline{\sigma }^{a}\tau ^{a}  \label{lss} \\
&&-\frac{\sigma ^{a}}{g\sqrt{\varsigma }}\left( \frac{g}{2}f^{abc}\overline{c%
}^{b}\overline{c}^{c}\right) +\frac{\sqrt{\varsigma }}{g}\sigma ^{a}L^{a}+%
\frac{1}{\varsigma }\left( \frac{g}{2}f^{abc}c^{b}c^{c}\right) \left( \frac{g%
}{2}f^{ade}\overline{c}^{d}\overline{c}^{e}\right) \;.  \nonumber
\end{eqnarray}
With the introduction of the auxiliary fields $\left( \sigma ^{a},\overline{%
\sigma }^{a}\,\right) $, the generating functional $\mathcal{W}(L,\tau )$
becomes
\begin{equation}
\exp i\mathcal{W}(L,\tau )=\int \left[ D\varphi \right] D\sigma D\overline{%
\sigma }\exp i\left[ S(A,\sigma ,\overline{\sigma })+\int d^{4}x\frac{\sqrt{%
\varsigma }}{g}\left( \sigma ^{a}L^{a}+\overline{\sigma }^{a}\tau
^{a}\right) \right] \,\;,  \label{wjs}
\end{equation}
with
\begin{eqnarray}
S(A,\sigma ,\overline{\sigma }) &=&S_{\mathrm{YM}}+S_{\mathrm{m}}+S_{\mathrm{%
gf}}+\int d^{4}x\left( -\frac{1}{g^{2}}\sigma ^{a}\overline{\sigma }^{a}\,+%
\frac{\overline{\sigma }^{a}}{g\sqrt{\varsigma }}\frac{g}{2}%
f^{abc}c^{b}c^{c}\right.   \nonumber \\
&&\,\,\,\,\,\,\,\,\,\,\,\,\,\,\left. \;\;\;-\frac{\sigma ^{a}}{g\sqrt{%
\varsigma }}\left( \frac{g}{2}f^{abc}\overline{c}^{b}\overline{c}^{c}\right)
+\;\frac{1}{\varsigma }\left( \frac{g}{2}f^{abc}c^{b}c^{c}\right) \left( 
\frac{g}{2}f^{ade}\overline{c}^{d}\overline{c}^{e}\right) \right) \;. 
\nonumber \\
&&  \label{sasgh}
\end{eqnarray}
As in the case of the gauge condensate $\left\langle A^{2}\right\rangle ,$
differentiating $\mathcal{W}(L,\tau )$ with respect to $L$ and $\tau ,$ one
obtains
\begin{eqnarray}
\left. \frac{\delta \mathcal{W}(L,\tau )}{\delta L^{a}}\right| _{L,\tau =0}
&=&\frac{\sqrt{\varsigma }}{g}\left\langle \sigma ^{a}\right\rangle
_{S(A,\sigma ,\overline{\sigma })}\;,  \label{dlts} \\
\left. \frac{\delta \mathcal{W}(L,\tau )}{\delta \tau ^{a}}\right| _{L,\tau
=0} &=&\frac{\sqrt{\varsigma }}{g}\left\langle \overline{\sigma }%
^{a}\right\rangle _{S(A,\sigma ,\overline{\sigma })}\;,  \nonumber
\end{eqnarray}
so that 
\begin{equation}
\left\langle \sigma ^{a}\right\rangle =\frac{g^{2}}{2\sqrt{\varsigma }}%
\left\langle f^{abc}c^{b}c^{c}\right\rangle
\;\;\;\;,\;\;\;\;\;\;\;\;\left\langle \overline{\sigma }^{a}\right\rangle =-%
\frac{g^{2}}{2\sqrt{\varsigma }}\left\langle f^{abc}\overline{c}^{b}%
\overline{c}^{c}\,\right\rangle \;.  \label{ssb}
\end{equation}
We see that the ghost condensates are related to the nonvanishing vacuum
expectation value of $\sigma ^{a},\overline{\sigma }^{a}$ evaluated with the
effective action $S(A,\sigma ,\overline{\sigma })$ of equation $\left( \ref
{sasgh}\right) .$ Let us proceed thus with the evaluation of the one-loop
effective potential for $\sigma ^{a},\overline{\sigma }^{a}$. To this order
it is sufficient to consider only the terms of the action $\left( \ref{sasgh}%
\right) $ which depend quadratically on the ghost fields $c^{a},\,\overline{c%
}^{a},$ namely
\begin{eqnarray}
S_{c\overline{c}}^{\mathrm{quad}} &=&\int d^{4}x\,\left( -\frac{1}{g^{2}}%
\sigma ^{a}\overline{\sigma }^{a}\;+\overline{c}^{a}\partial ^{2}c^{a}+\frac{%
\overline{\sigma }^{a}}{2\sqrt{\varsigma }}f^{abc}c^{b}c^{c}-\frac{\sigma
^{a}}{2\sqrt{\varsigma }}f^{abc}\overline{c}^{b}\overline{c}^{c}\right) \; 
\nonumber \\
&=&\int d^{4}x\left[ -\frac{1}{g^{2}}\sigma ^{a}\overline{\sigma }^{a}+\frac{%
1}{2}\left( 
\begin{tabular}{ll}
$\overline{c}^{a}$ & $c^{a}$%
\end{tabular}
\right) \mathcal{M}^{ab}\left( 
\begin{tabular}{l}
$\overline{c}^{b}$ \\ 
$c^{b}$%
\end{tabular}
\right) \right] \;,  \label{scc}
\end{eqnarray}
where $\mathcal{M}^{ab}$ denotes the $\left( N^{2}-1\right) \times \left(
N^{2}-1\right) $ matrix
\begin{equation}
\mathcal{M}^{ab}=\left( 
\begin{tabular}{ll}
$-\frac{1}{\sqrt{\varsigma }}\sigma ^{c}f^{cab}$ & $\partial ^{2}\delta
^{ab} $ \\ 
$-\partial ^{2}\delta ^{ab}$ & $\frac{1}{\sqrt{\varsigma }}\overline{\sigma }%
^{c}f^{cab}$%
\end{tabular}
\right) \;.  \label{mab}
\end{equation}
For the one-loop effective potential we get

\begin{equation}
V^{\mathrm{eff}}(\sigma ,\overline{\sigma })=\frac{1}{g^{2}}\sigma ^{a}%
\overline{\sigma }^{a}+\frac{i}{2}\mathrm{tr\log \det }\mathcal{M}^{ab},
\label{vss}
\end{equation}
where $\sigma ^{a},\overline{\sigma }^{a}$ have to be considered constant
fields. Let us focus on the case of $SU(2).$ In that case $%
f^{abc}=\varepsilon ^{abc}\;(\varepsilon ^{123}=1)$, and $\mathcal{M}^{ab}$
is a $6\times 6$ matrix. After a straightforward computation one has

\begin{equation}
V^{\mathrm{eff}}(\sigma ,\overline{\sigma })=\frac{1}{g^{2}}\sigma ^{a}%
\overline{\sigma }^{a}+i\int \frac{d^{d}k}{\left( 2\pi \right) ^{d}}\log
\left( \left( k^{2}\right) ^{2}+\frac{\sigma ^{a}\overline{\sigma }^{a}}{%
\varsigma }\right) \;.  \label{vss1}
\end{equation}
>From 
\begin{equation}
\int \frac{d^{d}k}{\left( 2\pi \right) ^{d}}\log \left( \left( k^{2}\right)
^{2}+\varphi ^{2}\right) =\frac{i}{32\pi ^{2}}\varphi ^{2}\left( -\frac{1}{%
\varepsilon }-2\gamma +2\log 4\pi -\log \frac{\varphi ^{2}}{\mu ^{4}}%
+3\right) \;,  \label{igh}
\end{equation}
the effective potential in the $\overline{MS}$ scheme is found to be

\begin{equation}
V^{\mathrm{eff}}(\sigma ,\overline{\sigma })=\frac{1}{g^{2}}\sigma ^{a}%
\overline{\sigma }^{a}+\frac{1}{32\pi ^{2}}\frac{1}{\varsigma _{0}}\sigma
^{a}\overline{\sigma }^{a}\left( \ln \frac{\sigma ^{a}\overline{\sigma }^{a}%
}{\varsigma _{0}\overline{\mu }^{4}}-3\right) \;.  \label{veffms}
\end{equation}
The minimum of the effective potential $\left( \ref{veffms}\right) $ is
given by the condition

\begin{equation}
\sigma _{\mathrm{\min }}^{a}\overline{\sigma }_{\mathrm{\min }%
}^{a}=\varsigma _{0}\overline{\mu }^{4}e^{2}\exp \left( -\frac{32\pi
^{2}\varsigma _{0}}{g^{2}}\right) \;,  \label{minms}
\end{equation}
which, up to one-loop order, is physically consistent for $\varsigma _{0}>0$%
. Setting

\begin{equation}
\sigma ^{a}=\sigma \delta ^{a3}\;,\;\;\;\;\overline{\sigma }^{a}=\overline{%
\sigma }\delta ^{a3}\;\;,  \label{3d}
\end{equation}
we obtain 
\begin{equation}
V^{\mathrm{eff}}(\sigma ,\overline{\sigma })=\frac{1}{g^{2}}\sigma \overline{%
\sigma }+\hbar \frac{1}{32\pi ^{2}}\frac{1}{\varsigma _{0}}\sigma \overline{%
\sigma }\left( \ln \frac{\sigma \overline{\sigma }}{\varsigma _{0}\overline{%
\mu }^{4}}-3\right) \;. \label{vefffmss}
\end{equation}
Let us now evaluate $\varsigma _{0}$. The first step is to obtain the
running of the fields $\sigma $, $\overline{\sigma }$. This task is easily
accomplished by recalling that, in the Landau gauge, the Faddeev-Popov
ghosts renormalize as \cite{landau,dsv} 
\begin{eqnarray}
c_{b} &=&Z_{c}^{1/2}c\;,  \label{ghr} \\
\overline{c}_{b} &=&Z_{c}^{1/2}\overline{c}  \nonumber
\end{eqnarray}
with 
\begin{equation}
Z_{c}=Z_{g}^{-1}Z_{A}^{-1/2}
\end{equation}
where

\begin{equation}
g_{b}=Z_{g}g\;,\;\;\;A_{b}=Z_{A}^{1/2}A\;.  \label{us}
\end{equation}
where the $g_{b},$ $A_{b},$ $\overline{c}_{b},$ $c_{b}$ denote the bare
coupling constant and fields. From the absence \cite{landau,dsv} of the
counterterms $L^{a}f^{abc}c^{b}c^{c}$, $\tau ^{a}f^{abc}\overline{c}^{b}%
\overline{c}^{c}$ it follows

\begin{eqnarray}
L_{b}g_{b}c_{b}^{2} &=&Lgc^{2}\;,  \label{ab} \\
\tau _{b}g_{b}\overline{c}_{b}^{2} &=&\tau g\overline{c}^{2}\;,  \nonumber
\end{eqnarray}
so that

\begin{eqnarray}
L_{b} &=&Z_{A}^{1/2}L\;,\;\;\;\;\gamma _{L}=\gamma _{A}  \label{rl} \\
\tau _{b} &=&Z_{A}^{1/2}L\;,\;\;\;\;\gamma _{\tau }=\gamma _{A}\;,  \nonumber
\end{eqnarray}
where $\gamma _{A}$ is the anomalous dimension of the gauge field $A_{\mu
}^{a}$. Therefore, making use of the equations $\left( \ref{ssb}\right) $,
for the running of $\sigma $, $\overline{\sigma }$ we find 
\begin{eqnarray}
\mu \partial _{\mu }\sigma &=&\hbar \left( \gamma _{A}^{(1)}+\frac{\beta
_{g}^{(1)}}{g}-\frac{1}{2}\frac{\beta _{g}^{(1)}}{\varsigma _{0}}\frac{%
\partial \varsigma _{0}}{\partial g}\right) \sigma +O(\hbar ^{2})\;,
\label{rssb} \\
\mu \partial _{\mu }\overline{\sigma } &=&\hbar \left( \gamma _{A}^{(1)}+%
\frac{\beta _{g}^{(1)}}{g}-\frac{1}{2}\frac{\beta _{g}^{(1)}}{\varsigma _{0}}%
\frac{\partial \varsigma _{0}}{\partial g}\right) \overline{\sigma }+O(\hbar
^{2})\;.  \nonumber
\end{eqnarray}
Requiring now the renormalization group invariance of the effective
potential $V^{\mathrm{eff}}(\sigma ,\overline{\sigma })$, namely 
\begin{equation}
\mu \frac{dV^{\mathrm{eff}}(\sigma ,\overline{\sigma })}{d\mu }=0+O(\hbar
^{2})\;,  \label{rgegh}
\end{equation}
for the coefficient $\varsigma _{0}$ one obtains

\begin{equation}
\frac{1}{8\pi ^{2}}\frac{1}{\varsigma _{0}}=\frac{2}{g^{2}}\gamma
_{A}^{(1)}\;.  \label{xiv}
\end{equation}
Finally, recalling that in the case of $SU(2)$ \cite{gr} 
\begin{eqnarray}
\beta _{g}^{(1)} &=&-\left( \frac{22}{3}-\frac{2}{3}n_{f}\right) \frac{g^{3}%
}{16\pi ^{2}}\;,  \label{ga1} \\
\gamma _{A}^{(1)} &=&\left( -\frac{13}{3}+\frac{2}{3}n_{f}\right) \frac{g^{2}%
}{16\pi ^{2}}\;,  \label{ga2}
\end{eqnarray}
it follows from $\left( \ref{ga2}\right) $ that

\begin{equation}
\frac{1}{\varsigma _{0}}=\left( \frac{2n_{f}-13}{3}\right) \;,  \label{fxi}
\end{equation}
completing therefore the evaluation of the one-loop effective potential for
the ghost condensates. A few remarks are now in order.

\begin{itemize}
\item  By combining the LCO technique with the BRST algebraic
renormalization we have been able to obtain the one-loop effective potential
for the ghost condensates. By construction, the effective potential $V^{%
\mathrm{eff}}(\sigma ,\overline{\sigma })$ obeys the renormalization group
equation.

\item  From the equations $\left( \ref{ga1}\right) $, $\left( \ref{ga2}%
\right) $ we observe that when the number of fermions $n_{f}$ is chosen so
that 
\begin{equation}
6<n_{f}<11\;,  \label{nf}
\end{equation}
we get a positive value for $\varsigma _{0}$, while keeping $\beta _{g}<0$,
so as to ensure asymptotic freedom. This result gives an evidence for the
existence of these condensates. We also observe that even for $n_{f}\leq 6$
the effective potential $V^{\mathrm{eff}}(\sigma ,\overline{\sigma })$ seems
to admit a nonvanishing minimum. However, in this case higher loops
contributions are needed in order to obtain a deeper understanding of that
minimum \cite{w}.

\item  In the present work we have considered the charged ghost polynomials $%
f^{abc}c^{b}c^{c}$ and $f^{abc}\overline{c}^{b}\overline{c}^{c}$. The same
procedure could have been applied to study the ghost polynomial $f^{abc}c^{b}%
\overline{c}^{c}$, which has vanishing Faddeev-Popov charge. As underlined
in \cite{work}, these operators correspond to different channels for the
ghost condensation, being related to the generators of the global symmetry $%
SL(2,R)$ present in the Landau gauge. It is worth mentioning that the
existence of different channels for the ghost condensation has an analogy in
superconductivity, known as the BCS\footnote{%
Particle-particle and hole-hole pairing.} versus the Overhauser\footnote{%
Particle-hole pairing.} effect \cite{ov}. In the present case the
Faddeev-Popov charged condensates $\left\langle
f^{abc}c^{b}c^{c}\right\rangle $, $\left\langle f^{abc}\overline{c}^{b}%
\overline{c}^{c}\right\rangle $ would correspond to the BCS channel, while $%
\left\langle f^{abc}c^{b}\overline{c}^{c}\right\rangle $ to the Overhauser
channel. Although our aim here is that of showing how to construct the
effective potential for the ghost condensation in the Landau gauge, we
remark that both channels could be studied simultaneously in order to decide
which one has the lowest energy \cite{w}.

\item  Finally, let us give the ghost propagators in the condensed vacuum,
namely
\begin{eqnarray}
\left\langle \overline{\,c}^{\alpha }(p)\;c^{\beta }(-p)\right\rangle  &=&-i%
\frac{p^{2}\,\,\delta ^{\alpha \beta }}{(p^{2})^{2}+v^{4}}%
\,\,\,\,\,,\;\;\,\;\;\;\;\;\alpha ,\beta =1,2\;  \nonumber \\
\left\langle \overline{\,c}^{3}(p)\;c^{3}(-p)\right\rangle  &=&-\frac{i}{%
p^{2}}\;.  \label{prop}
\end{eqnarray}

where 
\begin{equation}
v^{4}=\frac{\,\,\overline{\sigma }_{\mathrm{\min }}\sigma _{\mathrm{\min }}}{%
2\varsigma _{0}}  \label{cond}
\end{equation}

One see thus that the off-diagonal propagators get deeply modified in the
infrared region, the ghost condensation $\overline{\sigma }_{\mathrm{\min }}\sigma _{\mathrm{\min }}$ providing the infrared cutoff.
\end{itemize}

\section{Conclusion}

In this work the one-loop effective potential for the ghost condensates $%
\left\langle cc\right\rangle $ and $\left\langle \overline{c}\overline{c}%
\right\rangle \;$has been obtained in the Landau gauge by combining the
local composite operators  technique (LCO) with the BRST algebraic
renormalization. Our results might signal a deeper physical meaning of these
condensates, which have been proven to exist in other gauges \cite
{ms,k,sp,cf,cf1,cf2}. The existence of nonvanishing ghost condensates in the
Landau gauge can also be justified by symmetry breaking considerations.
Although the Landau gauge does not possess a quartic ghost self-interaction,
it shares an important feature with the nonlinear gauge considered in \cite
{cf,cf1,cf2}. Both gauges display indeed a global $SL(2,R)$ symmetry \cite
{sl2r}, present in $SU(N)$ Yang-Mills for any value of $N$. It is very
interesting to point out that the $SL(2,R)$ symmetry is also present in the
maximal Abelian gauge in the case of $SU(2)$, as underlined by \cite{ms}.
Recent investigations have proven that $SL(2,R)$ is indeed present in this
gauge for a generic $SU(N)$ \cite{work}. Therefore, the existence of ghost
condensates in these gauges can be seen as the manifestation of a very
general feature, \textit{i.e. }the dynamical symmetry breaking of $SL(2,R)$.

As one can see from the expressions $\left( \ref{prop}\right) $, the ghost
condensates modify the behavior of the propagator in the infrared region.
Recent lattice simulations for the propagator in the Landau gauge are nicely
reproduced by a form factor $F(p^{2})$ whose infrared behavior is given by 
\cite{lg}

\begin{equation}
\sum_{\mu a}\left\langle A_{\mu }^{a}(p)A_{\mu }^{a}(-p)\right\rangle =\frac{%
F(p^{2})}{p^{2}}\;,  \label{gprop}
\end{equation}
with

\begin{equation}
F(p^{2})=\mathcal{N}\frac{p^{2}}{p^{2}+m_{1}^{2}}\left[ \frac{1}{%
p^{4}+m_{2}^{4}}+\frac{s}{\left( \log \left( m_{L}^{2}+p^{2}\right) \right)
^{13/22}}\right] \;,  \label{fp}
\end{equation}
where $\mathcal{N},$ $s,\;m_{1},\;m_{2}\;$and $m_{L}\;$are fitting
parameters \cite{lg} whose values are $\mathcal{N}=8.113,\;s=0.32,%
\;m_{1}=0.64\;\mathrm{GeV},\;m_{2}=1.31\;\mathrm{GeV\;}$and $m_{L}=1.23\;%
\mathrm{GeV}$. Although a fully theoretical analytic derivation of this
behavior is still lacking, it is tempting to argue that both the gauge $%
\left\langle A^{2}\right\rangle \;$and the ghost condensates $\left\langle
cc\right\rangle ,\;\left\langle \overline{c}\overline{c}\right\rangle
,\;\left\langle c\overline{c}\right\rangle $ could give some useful insights
in order to improve our knowledge of the infrared region. Needless to say,
the gauge and the ghost propagators mix nontrivially in the Schwinger-Dyson
equations for the Landau gauge.

Finally, we point out that the LCO technique can be combined with the BRST\
algebraic renormalization to provide a powerful framework for analysing the
quantum properties of local composite operators. Particular attention will
be devoted to the condensate $\left( \frac{1}{2}\left\langle AA\right\rangle
-\xi \left\langle c\overline{c}\right\rangle \right) $ in the maximal
Abelian gauge. This condensate is BRST\ on-shell invariant and it is
expected to provide effective masses for both off-diagonal gauge and ghost
fields \cite{ope,dd}, a point of great relevance for the Abelian dominance. 

\section*{Acknowledgments}

The Conselho Nacional de Desenvolvimento Cient\'{i}fico e Tecnol\'{o}gico
CNPq-Brazil, the Funda{\c{c}}{\~{a}}o de Amparo {\ a Pesquisa do Estado do
Rio de Janeiro (Faperj) and the SR2-UERJ are acknowledged for the financial
support. We are grateful to D. Dudal and to M. Picariello for fruitful
discussion. }

\end{document}